# Developments and Obstacles in Chinese eBook Market

Author: SHEN Liang

# Abstract

The purpose of this study was to provide insights into the eBook market in China through case studies on eBook companies and a survey research with individual eBook users. The information from three companies, Beijing Superstar Electric Company, Beijing Founder APABI Technology Limited, and Beijing Sursen Electronic Technology Company Limited, showed that the B2B market has been developed due to the huge requirement from organization customers, universities libraries in particularly, and the B2C market is still immature. The information from interviews and relative data revealed that both Superstar and Sursen have serious copyright infringement which is an important problem impeding the further development of the eBook market. The questionnaire explored awareness, purchase, reading and other experiences of eBook end-users. Questions indicated that readers were attracted by the technical advantages including costless to copy, easy to transfer, searchable and easy to store, but did not want to pay for eBooks. Because the computers, especially desktop PCs, were the main device for reading and the CRT displays were massive used while there were few dedicated reading device in the market, many eBook end-users still preferred to read extended passages of text on papers rather than screens. Today the copyrights issue, user acceptance and the reading device are three significant obstacles for eBook industry in China.

# Keywords

Electronic Books, Electronic Publishing, Ebook Companies, Ebook User, Copyright, User Acceptance, Reading Device

# Contents

# Introduction

The hasty growth of information and communication technologies has dramatically driven the development of publishing. Electronic publishing has emerged as a fast growing phenomenon in the revolution of new media. The flourish of the Internet has showed its revolutionary impact to traditional publishing industry and also created many new opportunities by the innovating ways to spreading digital content to consumer markets. Electronic publishing can deliver information much faster and wider than traditional publishing and when the information technology infiltrates among consumers, it will possibly become the major force of publishing in the future. At the early years, the emergence of the eBook industry was accompanied by great hype and enthusiasm by industry experts, IT companies, publishing executives and many in the information field (Herther, 2005).

"EBooks promise to revolutionize the way the world reads. (Gates, 2001)"

" With sufficiently robust digital-rights management in place, libraries might even disappear, replaced by ftp or Web servers. (Dipert, 2003)"

However, neither scholars nor the government paid much attention to the electronic publishing industry in China until recent years. For understand the development of Chinese eBook market, this article provided insights into the history and status quo of eBook market in China through case studies on three biggest eBook companies and a survey research with individual eBook users.

The International Digital Publishing Forum (IDPF), formerly the Open eBook Forum (OeBF) is an international nonprofit trade organization whose mission is to improve the development of the electronic publishing market. In 2000, the Open eBook Forum published *A Framework for the Epublishing Ecology*, which provides a systematic foundation for critical thinking and discussion in the world of electronic publishing. The framework is composed of three parts: Glossary, Reference Model, and Stakeholder Profiles.

In *Glossary*, the most important terms including eBook have been thoroughly reviewed and precisely defined. Open eBook Forum defined electronic book as:

1. A literary work in the form of a Digital Object, consisting of one or more standard unique identifiers, Metadata, and a Monographic body of Content, intended to be published and accessed electronically. 2. May also refer to the hardware device created for the purpose of reading eBooks. (OeBF, 2000, p 6)

The former can be regarded as *soft eBook* and the latter means *hard eBook*. Generally speaking, people's first reaction to the word eBook is the *soft eBook* in China today due to the comparatively slower development of reading devices. In this article, the term *eBook* means the *soft eBook* as well.

The *Reference Model* and *Stakeholder Profiles* in *A Framework for the Epublishing Ecology* provided a good theoretical frame to analyze the interaction between roles and the requirements of a specific stakeholder in the market of electronic publishing. So the design of this study was mainly followed the International Digital Publishing Forum's theoretical frame to observed the eBook market in China from the perspective of technology, law and society.

In *Reference Model*, the epublishing world was regarded to be composed of multiple domains with four types of elements: Roles, Interactions, Objects and Authorities (OeBF, 2000, p 9). *Roles* are the "live" staff performed (or "played") by persons, organizations and systems (can be computer systems or software agents) (OeBF, 2000, p 12). The roles in the epublishing ecology are classified into three abstract categories: Originators, Intermediaries, and End-User (OeBF, 2000, p 10). *Originators* are people or organizations that create electronic publications, like authors, anthologists, and word processors; *Intermediaries* are people or organizations, editors, publishers, and agents for instance, who facilitate the flow of content and payments from *Originators* to *End-Users*; *End-Users* are the consumers who purchase electronic publications like individual readers, and library patrons (OeBF, 2000, p 10). *Objects* are the "dead" things, including rights objects, products, services, monetary objects, and so on, that flow between players of Roles. *Interactions* are what happen between players of roles (OeBF, 2000, p 12). *Authorities* are the rules like technical standards, government administration and social norms in the epublishing ecology, and they provide the governance context in which *Interactions* occur (OeBF, 2000, p 12).

In the epublishing ecology of China, there are three roles as well. *Originators* are mainly played by eBook companies, traditional publishers and authors. Most *Intermediaries* are performed by eBook companies who sell eBooks to consumers. Libraries and the individual eBook readers who buy eBooks from eBook companies play the major roles of *end-users*. Normally, traditional publishers release paper books first and generated digital versions of some ones after the sale periods are over and just a few of publishers directly sell eBooks to end-users, say play the role of intermediaries. Generally speaking, traditional publishers just sell the copyright authorizations of digital copies to eBook companies and let them convert the literary works into eBooks. Then these companies accomplished the production of eBooks and sell them by themselves. So eBook companies perform both *originators* and *intermediaries* at the same time.

In *Stakeholder Profiles*, stakeholders were defined as "entities (person, organization, etc.) with financial, economic, or moral interest in the epublication or the performance of one or more functions in the epublishing ecology" and divided into 8 broad categories: Originator, Rights Holder, Publisher, Service Provider, Technology provider, Seller, Distributor, and End-user (OeBF, 2000, p 17). An *Originator* is the author, editor or creator who creates or arranges for the creation of the intellectual property that will become the content of the electronic publication[1] (OeBF, 2000, p 17). A *rights holder* is an entity, generally the author or the publisher, that owns or has been licensed the digital rights for the intellectual property created by the originator (OeBF, 2000, p 18). *Publisher* is an entity that derives the creation of Literary Works[2] and prepares, promotes, and distributes them to wholesalers, retailers, or end-users (OeBF, 2000, p 18). *Service Provider* is an entity that provides an accessory service such as assisting in the creation, distribution or protection of the electronic publication or the collection and distribution of consumer information (OeBF, 2000, p 19). A *technology provider* is an entity that enables the electronic distribution of content by providing either software or hardware (OeBF, 2000, p 19). In fact, the line between *Service Provider Seller* and *Technology Provider* is very vague because of the integration of electronic publishing system. Today, these two stakeholders are normally only one entity: the eBook company. A seller is an entity that attracts consumers, enables them to browse and search metadata and ultimately sells the electronic publication to the consumer (OeBF, 2000, p 20). A *Distributor* is an entity that provides the epublication directly to an end-user or another distributor through a protected transaction (OeBF, 2000, p 20). In the epublishing ecology of China, eBook companies are not only *service providers* and *technology providers* but also *sellers* and *distributors* at the same time. They are the stakeholders who have the biggest economic interest from epublications. Therefore, eBook companies are particularly significant research objects in this article. An *end-user* is an eBook consumers (can be individual or organization) that purchases and/or accesses the epublication[1] (OeBF, 2000, p 20). End-users are as important as, if not more important than, eBook companies, because they determine the main requirement of the eBook market and the massive individual readers will point out the future of eBook. In this article, end-users, individual eBook readers in particular, are another important research objects.

---

1  In *A Framework for the Epublishing Ecology*, the term originator was used for either a *role* or a *shareholder*, and the two originators have different meanings.

2  From the U.S. Copyright Act 1976: "*Literary Works* are works, other than audiovisual works, expressed in words, numbers, or other verbal or numerical symbols or indicia, regardless of the nature of the material objects, such as books, periodicals, manuscripts, phonorecords, film, tapes, disks, or cards, in which they are embodied.

For the research on eBook companies, three biggest eBook companies, Beijing Founder APABI Technology Limited, Beijing Superstar Electric Company and Beijing Sursen Electronic Technology Company Limited, who shared the main eBook market, were selected for case studies. Data about them was mainly collected from interviews to people who are working or have worked in the field of epublishing in China. They provided many valuable first hand information including negative statements about eBook companies. So the participants' identity will be treated with confidentiality and not be disclosed in this article. In addition, my interviewees also provided important background information about the history of Chinese epublishing industry. Data about eBook companies was also collected from documents and reports from these companies, their costumers and relative conferences.

---

1  In *A Framework for the Epublishing Ecology*, the term *end-user* was used for either a *role* or a *shareholder*, but the two *end-user* was refer to the same entity.

Data about individual eBook users were collected from an online survey. A questionnaire (see Appendix) with nine questions was distributed to Internet users through several big online forums in September 2006. The questions explored awareness, purchase, reading and other experiences of eBook end-users. The online survey attracted 320 responses in total and they formed the base for subsequent analyses reported in this paper. The data was analyzed quantitatively as well as qualitatively. Considering the earlier development of eBook market in the United States and Europe, the data from end-users was also comparatively analyzed with similar studies in western countries.

# A Brief History of Chinese EBook Industry

## The Primitive Stage

1995 can be regarded as the beginning of electronic publishing in China. In that year the government launched the China Academic Journals (CD-ROM Version) project, which aimed at converting academic journals into electronic journals. Tsinghua Tongfang Knowledge Infrastructure Technology Company Limited got the contract from the government and became the biggest online database provider later. Soon the Ministry of Science and Technology of the People's Republic of China also decided to digitize its academic journals. Beijing Wanfang Data Company Limited and Chongqing VIP Information Company Limited got great developments during this project and all became important online ejournal database providers later. Today these three companies almost have occupied the whole cake of ejournal market in China.

The eBook industry debuted during the later years of the dotcom era in the west, a time when vendor capital was readily available and hopes for Internet-based business were very high (Herther, 2005). The same phenomena also took placed in China from 1998 to the earlier 2000, the early years of Chinese eBook market. In 1998, National Library of China began building National Digital Library of China and the government invested for this project. Superstar, one of the biggest eBook companies today, get the contract for converting 200,000 titles of paper books owned by National Library of China into digital format. Since 1999, eBook retail websites following the first series of e-commerce websites bloomed. In 1999, people.com made out a business-to-consumer (B2C) platform for eBook business and got contract with more than 200 publishers. Another B2C eBook retail website bookuu.com was also established in 1999 by some students who had studied in foreign countries and came back to start their owned business. Bookuu.com.cn has signed contract with more than 100 authors and have more than 1,000 eBooks in store. In 2000, Beijing Sursen Electronic Technology Company Limited, whose main field was digital documents, went into the eBook market of and began selling eBooks on its website. The period from 1995 to 1999 was the primitive stage of the electronic publishing industry in China in which the ecology of epublishing emerged and rapidly grew.

## The Trial And Error Phase

In 2000, the winter for dotcom befell in both east and west, and the bobbles of many dotcoms busted. Companies only based on new conceptions of like eBooks can neither become profitable nor attract vendor capitals that have become more realistic. In this year, Bookoo.com.cn announced their decision to end the e-book business in June 2001 (Kang, 2001). People.com also eventually quitted the eBook market in 2000. Similar condition for the eBook market in the United States also occurred In 2001. Newly created eBook divisions within large publishing companies like Random House were closed and netLibarary, a famous eBook provider, went into free fall (Hawkins, 2002). During this period, not only in China bout also in the world, eBook companies were seeking a profitable business model and the direction for further development. So the period from 2000 to 2001 can be regarded as the trial and error phase for the electronic publishing industry.

## The Developing Period

Facing with the difficulty of B2C model in electronic publishing market, eBook companies changed their strategy and turned to the business-to-business (B2B) mode. EBook companies begin to focus on eBook wholesale to organization consumers like libraries for instead of retailing eBooks to individual consumers. From 2002 to 2004, B2B model became the main model for eBook companies, and that made these companies start profiting. In the period, the customers for eBook were mainly libraries in universities. Renmin University of China, for instance, paid 6,600,000 CNY to added books 2003, and 27.7 per cent of the expenditure was used for eBooks, which were 41 per cent of the whole purchased books. Beijing University of Posts and Telecommunications in 2003 paid 1,700,000 CNY, which was about half of its expenditure, for the construction of its digital library.

The success of B2B model in the eBook market did not only enable eBook companies to survive through the winter but also got substantial growth. According to the data from *2004-2005 Annual Report of Publishing Industry in China*, 8,050,000 eBooks were sold out in 2004 and this number was 2.6 times of 2003. Until May 2005, there have been 148,000 titles of eBook published and about one hundred traditional publishers have started their epublishing business (Hao, 2005). There were 26 publishers got more than 300,000 CNY total revenue from eBook, 25 publishers got more than 500,000 CNY, and 5 publishers got more than 1,000,000 CNY (Ren, 2005). In recent years, the infrastructure for IT industry has been improved, and the environment was more mature for e-commerce. Because more and more people were getting used to read online, the high hope for the B2C market was reignited. This era from 2002 to today was the developing period of the electronic publishing industry and this period will be last for long time.

# EBook Companies

## Beijing Superstar Electric Company

Beijing Superstar Electric Company is a private enterprise established in 1993 and also one of the oldest eBook companies in China. At its early years, Superstar's main business was converting paper documents into digital files by using its PDG format for the government and academies. During this period, Superstar was just a service and technology provider who had limited interest in the epublishing word, and only played the role as an Originator.

In 1996, Superstar ventured into the electronic publishing business and produced over 200 products, mainly in the media of CD-ROM, including some older periodicals and historical archival documents. At the end of 1997, Superstar developed a web-based eBook publishing system. In 1998, National Library of China (NLC) began to build National Digital Library of China, and the government invested 1,235,000,000 CNY for this project. Superstar got the contract and established a partnership with the National Library of China. During July 1998 to December 1999, Superstar converted 200,000 titles of paper books owned by NLC into electronic book. This was generally regarded as the one of the most successful case of eBook in the early market and Superstar was extremely well received. According to the information from an interviewee who has worked in Superstar, Superstar earned approximate one million CNY.

However Superstar did not only earn a lot of money from this project but also kept the copies of those 200,000 titles of eBooks, which was extremely important for the later development of Superstar. In 2000, Superstar began establishing its digital library and the eBooks derived from NLC formed the main part of the Superstar Digital Library. Since 2000, Superstar started to sell these books and digitized more paper books and the huge amount of eBooks made Superstar more competitive than other companies.

In Chinese eBook market, end-users were played by two groups: organization consumers and individual consumers, and they are all the target customers of Superstar. Libraries, especially universities libraries, were first drawn by the digital library owing to the education policy of the government. Project 211, a constructive project for universities and colleges conducted by the government of the People's Republic of China since 1995, tremendously promoted the requirement of eBooks in universities' libraries. In Project 211, about a hundred universities were selected and subsidized for upgrading and improvement of their infrastructure for teaching and research ("a Brief Introduction of Project 211"). For the candidate universities, the amount of books and documents in their libraries was an important benchmark for the government's examination. Soon the book number became a common benchmark for to government to measure universities' quality. To attract the financial aid and deal with the examinations from the government, universities started to buy more books for their libraries and eBooks became their first choice because eBooks were much cheaper than traditional books. The average price for an eBook was only about 1 CNY while a paper book cost ten times and maybe more, and a server costing less than 100,000 CNY can store more than 200,000 titles of eBooks. According to an interviewee's information, Superstar got approximate 300 organization customers, mainly university libraries, in three years. Today most Chinese universities have bought the mirror sites of Superstar Digital Library.

For individual users, Superstar's strategy was different from normally eBook retailers like Founder APABI and Sursen, and adapted its unique library mode: lending eBooks to individual users rather than selling. Like traditional libraries, users of Superstar Digital Library have to pay for reader cards to download eBooks and use Superstar's own reading software to read the eBooks in PDG format. EBooks borrowed from Superstar Digital Library can be only read in limited days and will be delete automatically after the deadline. It costs 15 CNY for a reader card for a month, 35 CNY for one season, 100 CNY for a year, and 188 CNY for two years[1]. According to an interviewee's data, there are more than 100,000 registered members of Superstar Digital Library today. Now Superstar claims to have over 150,000 titles of eBooks for paid member of its digital library and offers 5,000 titles of eBook free for trail reading

---

1  Superstar Digital Library homepage http://www.ssreader.com/ssreadercard/ (accessed October 2005)

per day[1].

However the purchasing power of individual consumer still cannot compare with organization consumers today and B2C is still the main model of Superstar as all the eBook companies. The competition among these companies is mainly focused on the amount of eBook titles owing to the demand from B2B market. Superstar Digital Library is generally regarded as the biggest digital libraries in China, but the amount of its eBooks is a mystery. At the beginning, the amounts of eBooks in Superstar Digital Library can be easily found in its homepage, and the number were keeping increasing. However, in recent years the number seems stopped at 150,000. According to a document presented in the 2004 conference of Beijing College Net Library[2], Superstar Digital Library Develop*ment Report* by Wang Dong, an employer from the market department of Superstar, the real amount of eBooks held by Superstar was much larger than what it announced and keep increasing in a prodigious speed (Figure 1). The information from China Geological Library, a Superstar's customer who bought a mirror site of Superstar Digital Library, revealed that Superstar has owned at least 1,500,000 titles of eBook until October 2006[3]. This number is ten times of what Superstar has announced.

---

[1] Superstar Digital Library homepage http://www.ssreader.com/ (accessed October 2005)

[2] This document can be found at the homepage of Beijing Universities Net Library http://www.netlib.edu.cn/hyjy/ppt/%B3%AC%D0%C7.ppt (accessed October 2005)

[3] China Geological Library homepage http://www.cgs.gov.cn/NEWS/Geology%20News/2006/20061102/20061102003.htm (accessed October 2005)

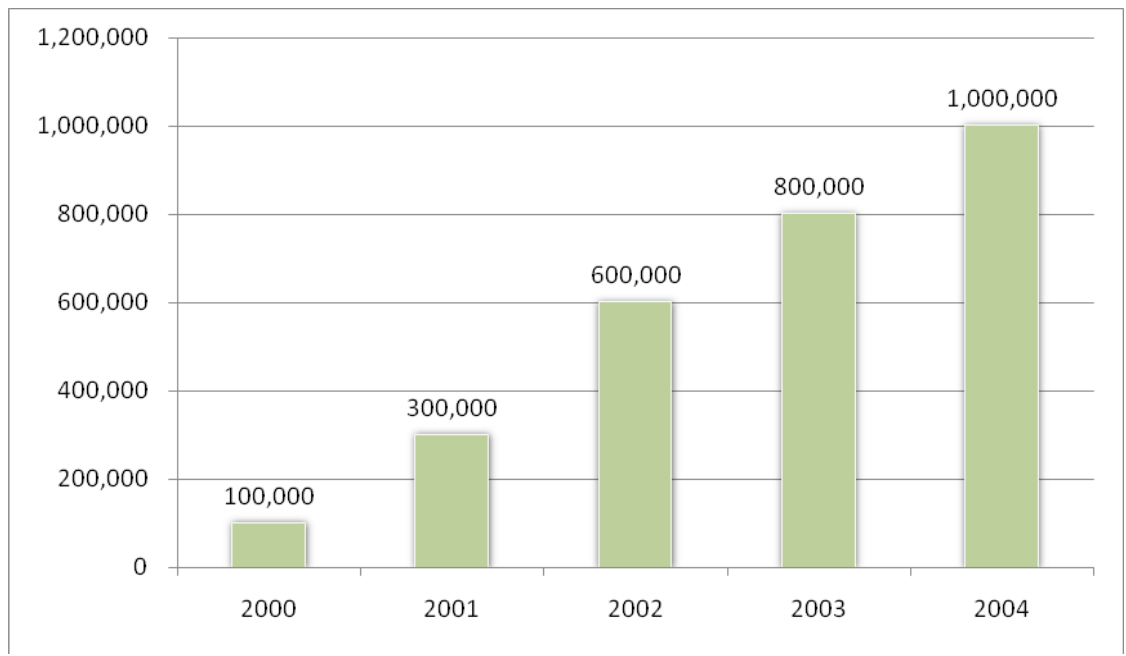

Figure 1 The EBook Number of Superstar Digital Library (Source: Beijing Superstar Electric Company 2004)

The reason for Superstar to hide its eBook numbers from the public is the copyright issue which eternally obsesses the electronic publishing in China. Since Superstar began to sell the 200,000 copies of eBooks derived from National Library of China, these books' rights holders were hardly considered at all. Although Superstar have signed contracts with 480 publishers, the main sources of the content of its eBooks were the paper books, which were digitized without authorization, from traditional publishing market. Superstar explained its behaviors by announcing that it have got the authorizations from the authors. Superstar Digital Library declared that it has got 253,253 authors' authorization until 31 April 2006[1]. If we use 2,000 CNY, the lowest price for one author's authorization, as the average cost to calculate, the total money for 253,253 authors' authorization will be more than five hundred million CNY which is impossible for Superstar to pay. An interviewee revealed that Superstar has gotten just authorizations from 15 per cent to 20 per cent of the whole authors of its eBooks. For selling eBooks without authorization from rights holders, Beijing Superstar Electric Company was sued for copyright infringement by Encyclopedia of China Publishing House in 2002 (Encyclopedia of China Publishing House v. Beijing Superstar Electric Company, 2002), by an author Li Haiwen (李海文) in 2005 (Li

---

[1] Superstar Digital Library homepage http://www.ssreader.com/zhuanti/15/ (accessed October 2005)

Haiwen v. Beijing Superstar Electric Company, 2005), by Cultural Relics Publishing House in 2005 (Cultural Relics Publishing House v. Beijing Sursen Electronic Technology Company Limited, 2005), by Beijing Normal University Press in 2006 (Beijing Normal University Press v. Beijing Superstar Electric Company, 2006), by eight authors Li Zheying (李哲英) (Li Zheying v. Beijing Superstar Electric Company, 2006), Liu Heping (刘和平) (Liu Heping v. Beijing Superstar Electric Company, 2006), Wang Xiaoming (王晓明) (Wang Xiaoming v. Beijing Superstar Electric Company, 2006), Zhang Yafei (张亚飞) (Zhang Yafei v. Beijing Superstar Electric Company, 2006), Zhuang Yongping (庄永平) (Zhuang Yongping v. Beijing Superstar Electric Company, 2006), Shen Hongxin (沈鸿鑫) (Shen Hongxin v. Beijing Superstar Electric Company, 2006), Zheng Hong (郑红) and Wu Guan (吴冠) (Zheng Hong and etc. v. Beijing Superstar Electric Company, 2006) in 2006, and by Social Sciences Academic Press (Social Sciences Academic Press v. Beijing Superstar Electric Company, 2006). Most of these lawsuits ended with imparlance and Superstar paid compensation to the authors. However these publishers and authors are the minority of the rights holders. If all of the rights holders assert their copyrights with the lawsuits, Superstar will definitely bankrupt for paying the huge compensation.

# Beijing Founder APABI Technology Limited

Beijing Founder APABI Technology Limited is a subsidiary of Founder Electronics Company Limited, which is one of the backbone subsidiaries of the Founder Group. Founded in 1986, Founder Group was a high-tech IT company originated Peking University, and now has become a multi-faceted group with several competitive IT subsidiaries. In China, Founder Group today owns 4 listed companies: Founder Holdings Limited, Founder Technology Group Corporation, EC-Founder (Holding) Company Limited, and Founder Electronics Company Limited.

Beijing Founder Electronics Co., Ltd. provides information processing technology, software, integrated solutions and value-added services for six main aspects: electronic publishing, digital media, electronic government, graphic software, IT equipment, network circulation[1]. Typesetting and publishing system is the earliest and most important business of Founder. Founder Electronics has accumulated more than 30 years experience in the publishing industry and has become the largest provider of publishing system. Founder's strategy for development is to expand from electronic typesetting to network typesetting, from single media to multimedia and cross-media, from the domestic market to the global market, from software business to software and services, and from Chinese information processing to multilingual information transmission. Electronic publishing is regarded as a new business growth area by Founder Electronics and will play an important role in the future.

During 30 August to 3rd September in 2000, the Eighth Beijing International Book Fair, Founder Electronics initially introduced its Internet publishing integrated solution APABI system[2]. The name of APABI was composted by the acronym of "Author", "Publisher", "Artery", "Buyer", and "Internet". APABI system contained five correspondent components: APABI Reader, APABI Maker, APABI Writer, APABI Rights Server, and APABI Retail Server. In 2001 Founder Electronics formally launched its APABI eBook integrated solution. Founder APABI is a comparatively new competitor, but with a much stronger IT background than other eBook companies. Founder's typesetting system is being used by over two thirds of publishers in China. This gives publishers the advantage of obtaining electronic copies of books virtually just with one click. This eBook system is packaged with an electronic resources management component that can be used by libraries and an online bookshop component that can be used by publishers to offer e-books directly to consumers. Besides, Founder provides low price for Founder APABI system to publishers who adapted its typesetting system. Founder APABI enable traditional publishers to play the roles of both originator and distributers and Founder APABI just gets profit as service provider and technology provider. That gives traditional publishing houses the opportunity to start their owned electronic publishing. Guangxi Normal University Press and the Juvenile & Children's Publishing House, for instance, have begun

---

[1] Beijing Founder Electronics Co., Ltd homepage http://www.founder.com.cn/en/AboutUs/index.asp?id=45 (accessed October 2005)

[2] Beijing Founder APABI Technology Limited homepage http://intro.APABI.com/news/new_1/index.htm (accessed October 2005)

offering the electronic versions of their paper books online via the Founder APABI system. By these strategies, Founder APABI gets what is more important than money: its CEB format is widely adapted in the epublishing industry and become one of the most popular eBook formats in China today.

Data from a presentation by Founder APABI in the 2004 Conference of Beijing College Net Library showed that more and more publishers have became partners with Founder APABI and the amount of eBook in APABI format increased dramatically (Table 1)[1]. In 2005, Founder APABI has got 400 contracted publishers, more than 20 partner eBook retailers, and 1,300 library subscribers including 40 overseas libraries, and 450 university subscribers. Until March 2006, the total number of eBooks owned by Founder APABI has reached 210,000 titles ("2005 Annual Report", 2005).

Table 1: The Partner Publishers and Published EBooks of Founder APABI

| Year | Partner Publishers | Published EBooks |
|---|---|---|
| 2000 | 16 | 49 |
| 2001 | 100 | 2,500 |
| 2002 | 200 | 20,200 |
| 2003 | 300 | 100,000 |
| Source: Beijing Founder APABI Technology Limited (2004) | | |

Soon after the launching of APABI integrated solution. Founder APABI also started to build its B2C platform www.apabi.com for eBook retail and a book search platform www.esoushu.com. However, because Founder APABI was not so successful in the B2C market as in the B2B market like most eBook company, the two web sites were merged into one later.

Because of the close relationship between Beijing Founder Electronics Co., Ltd. and Chinese publishers, Founder APABI can directly get the eBook licenses from the publishers and that gives more advantages to Founder APABI on the copyright issue than other eBook companies. An interviewee revealed that Founder APABI normally paid 40 per cent of the profit for the sale of each eBook to the publisher who holds the copyright. Founder APABI is considered as the only eBook company in China that has completely cleared copyright issues with right holders for its eBook collections.

---

1 This document can be found at the homepage of Beijing Universities Net Library
http://www.netlib.edu.cn/hyjy/ppt/%B7%BD%D5%FD.ppt (accessed October 2005)

Founder APABI was one of few Chinese companies involved in hardware development for hand held eBook readers. Founder APABI has developed several devices compatible with its own eBook format. When I interviewed an ex-employer of Founder I was told that even Founder APABI did not take its reading devices serious today, but it would keep researching and developing to wait for the maturity of both eBook market and hardware technology. By comparison, neither Superstar nor Sursen have enough technique and money to venture for the reading devices as Founder APABI whose holding company is one of the biggest computer hardware manufacturers in China.

## Beijing Sursen Electronic Technology Company Limited

Beijing Sursen Electronic Technology Company Limited was founded in 1996, and has more than 300 staffs including about 100 software engineers today. Sursen has been specializing in providing the technologies and solutions for replacing traditional paper files with its SEP format digital documents. At the very beginning, Sursen also has product some electronic publications for newspapers and journals. Sursen formally entered the market of electronic publishing in 2000 and started to play the roles of both Originators and Intermediaries. Sursen used its technique on digital documents to digitize paper books and convert into SEP format. Sursen's SEP format is based on XML and its full text can be searched. Sursen also developed an electronic resources management system for digital libraries and electronic books product system for publishers based on its owned digital documents management technique.

Like most eBook companies, Sursen aimed at both organization customers and individual customers and mainly focused on the B2B market. Sursen also harvest its main customers from education market promoted by Project 211. About 50 per cent of these universities, including Peking University and Tsinghua University, in Project 211 has became Sursen's customers. According the data revealed by Zhou Heping, a manager in Sursen, in the 2004 conference of Beijing College Net Library, Sursen has got 108 customers who bought its database pack and 520 customers who bought its mirror sites[1]. They are either university libraries or public libraries and the universities libraries are the great majority among them.

In 2000 Sursen also launch its online eBook retail platform www.21dmedia.com. However the B2C market cannot compare with the B2B market. The data from my survey on eBook readers indicated that the share of eBooks in Sursen's format in the B2C market was the lowest in these three eBook companies. According to the information from a former employer of Sursen, the total revenue of Sursen was approximate 100,000,000 CNY in 2003, but just a small part of them was come from its eBook business.

Compared with Superstar, Sursen enjoys a better reputation because of its identity of a high tech company and comparatively small market share. In fact, Sursen has done the same "dirty jobs" of copyright infringement like Superstar. Sursen has stopped published the amount of its eBook to the public for long time. The Library of China University of Mining and Technology, a customer of Sursen, revealed that there have been about 600,000 titles of eBooks in the digital library of Sursen until October 2006[2]. Although there have been about 400 publishers cooperated with Sursen today, Sursen do not have the close relationship with publishers like Founder APABI and still has difficulty to get rights holders' authorization. In fact, Sursen's main book source is the unauthorized paper books as well. Like Superstar, Sursen also ignored the rights holders during the pursuit of the amount of eBook titles as well. Beijing Sursen Electronic Technology Co., Ltd. was sued for copyright infringement by seven authors Zhen Chengsi （郑成思）(Zheng Sicheng v. Beijing Sursen Electronic Technology Company Limited, 2004), Li Shunde (李顺德) (Li Shunde v. Beijing Sursen Electronic Technology Company Limited, 2004), Tang Guangliang （唐广良）(Tang Guangliang v. Beijing Sursen Electronic Technology Company Limited, 2004), Zhang

---

1  This document can be found at the homepage of Beijing Universities Net Library
http://www.netlib.edu.cn/hyjy/ppt/%CA%E9%C9%FA.ppt (accessed October 2005)

2  Library of China University of Mining and Technology homepage (16 October 2006)
http://lib.cumt.edu.cn/temp/ss.txt (accessed October 2005)

Yurui (张玉瑞) (Zhang Yurui v. Beijing Sursen Electronic Technology Company Limited, 2004), Xu Jiali (徐家力) (Xu Jiali v. Beijing Sursen Electronic Technology Company Limited, 2004), Zhou Lin (周林) (Zhou Lin v. Beijing Sursen Electronic Technology Company Limited, 2004), Li Mingde (李明德) (Li Mingde v. Beijing Sursen Electronic Technology Company Limited, 2004) and Encyclopedia of China Publishing House (Encyclopedia of China Publishing House v. Beijing Sursen Electronic Technology Company Limited, 2004) in 2004, by five authors He Haiqun (何海群), He Huwei (何湖苇), Tang Ying (唐颖) (He Haiqun and etc. v. Beijing Sursen Electronic Technology Company Limited, 2005), Tie Zhuwei (铁竹伟) (Tie Zhuwei v. Beijing Sursen Electronic Technology Company Limited, 2005), Jin Chunming (金春明) (Jin Chunming v. Beijing Sursen Electronic Technology Company Limited, 2005) in 2005.

# Individual EBook Users

Since 2000, the internet and computer have been greatly developed in China and the amount of Internet users also increased in a staggering speed. Until 31 December 2005, there have been 111,000,000 internet users in China, and this number increased 393.3 per cent compared with the number from 2000 (Figure 2).

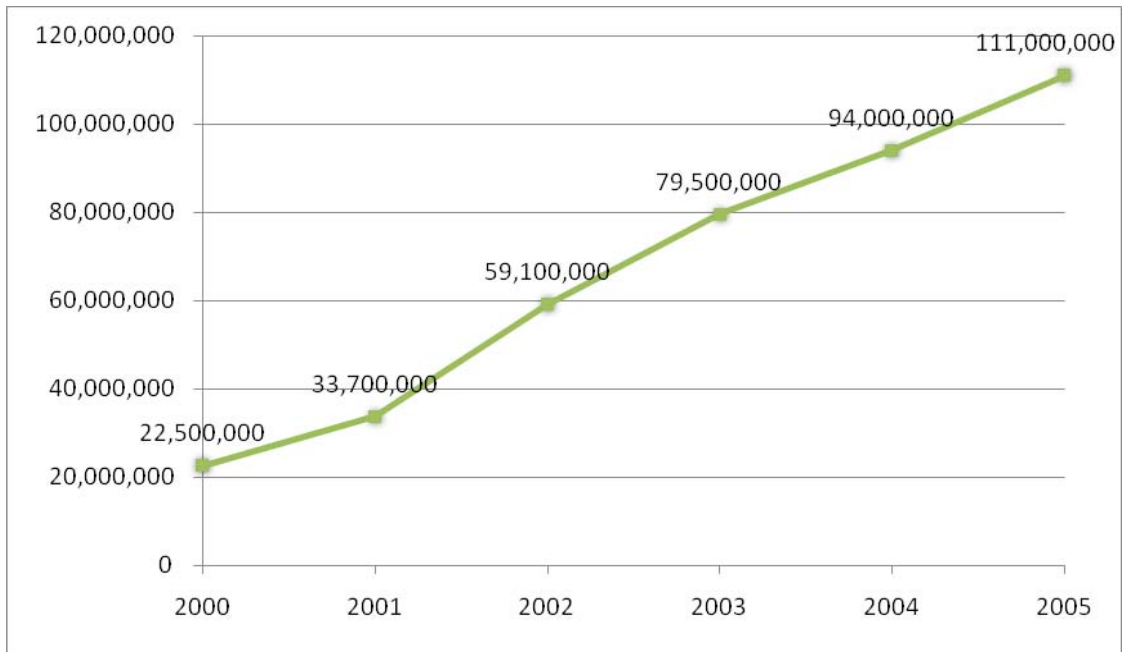

Figure 2: The Numbers of Internet Users in China (Source: CNNIC 2006)

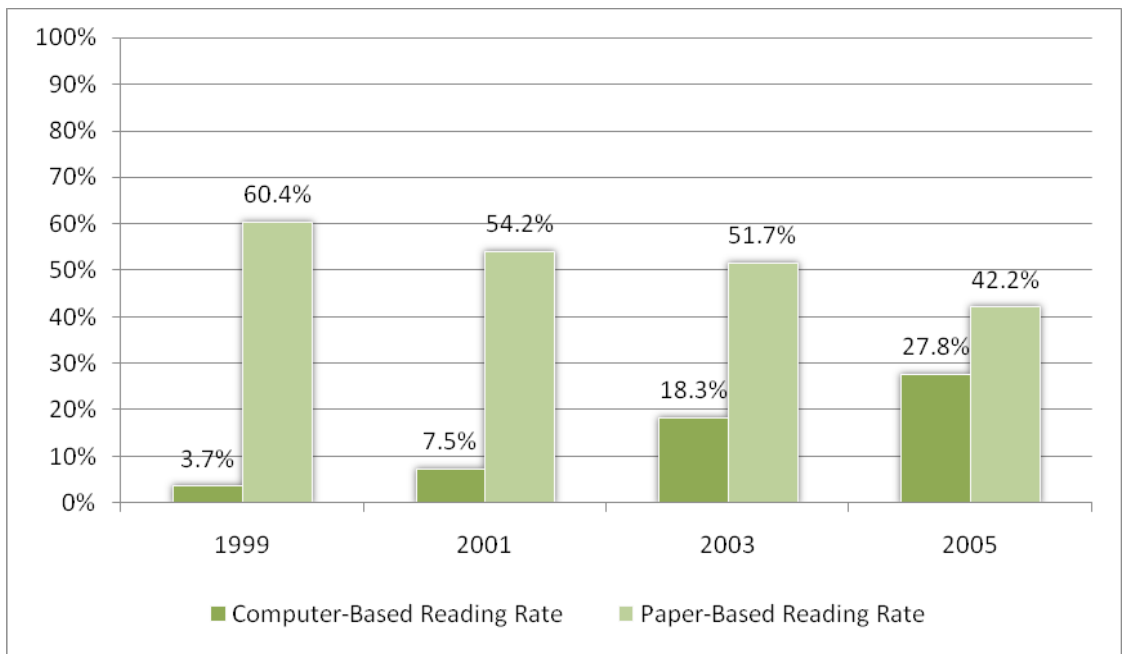

Figure 3: The Changing of People's Reading in China (Source: CIPS 2006)

However, according to the surveys of China Internet Network Information Center from 2001 to 2005, the electronic book has being regarded as the most dissatisfied aspect in the internet services by users (Table 2). This data was widely quoted by eBook companies to show the potential of Chinese eBook market in China, but the real statue was far from that. The fail of B2C model has showed that simply statistics may be indicative of market potential, but without knowing the eBook consumers' reading experience, eBook companies may lack information that can show the obstacles in the B2C market. Therefore this study represents an attempt to seek feedback directly from Internet users about their eBook experiences.

Table 2: eBook Readers and Dissatisfaction with eBook in Chinese Internet Users

| Year | EBook Readers | Dissatisfied with eBook |
|---|---|---|
| 2001-01 | 45.99% | 39.34% |
| 2001-07 | 32.80% | 29.10% |
| 2002-01 | 37.40% | 36.30% |
| 2002-07 | 35.60% | 37.70% |
| 2003-01 | 32.60% | 38.90% |
| 2003-07 | 31.50% | 36.80% |
| 2004-01 | 28.80% | 31.10% |
| 2004-07 | 27.50% | 28.80% |
| 2005-01 | 36.70% | 31.08% |
| 2005-07 | 33.30% | 25.10% |
| Source: CNNIC (2001-2005) | | |

## Ever Accessed eBooks

Questions about experiences on eBooks began from asking respondents whether they had ever read any eBook. There was an almost even split between respondents who answered yes (46.88%) and no (53.13%) to this question (Table 3). Of respondents who indicated that they had read eBooks, further questions were asked.

Table 3: Question 1: Have you read any eBook in the past?

| Frequency | Percentage | Option |
| 150 | 46.88% | Yes |
| 170 | 53.13% | No |

## The Source of eBooks

All respondents who had read eBooks were asked where they got their eBooks. Nearly all the respondents (96.47%) have gotten eBooks from websites that provide free eBooks, 20.59 percent readers have accessed eBooks from Websites that sale eBooks or charge in other ways, and few (3.53%) got eBook in CD-ROM format (Table 4). This data indicated that internet has become the main media for eBooks' transfer, and other digital media like CD-ROM has become "traditional" and almost has been replaced by the internet.

Table 4: Question 2: Where did you get eBooks?

| Frequency | Percentage | Option |
|---|---|---|
| 164 | 96.47% | Websites that provide free eBooks to download or browse |
| 35 | 20.59% | Websites that sale eBooks or charge in other ways |
| 52 | 30.59% | Websites of public libraries or libraries in universities or schools |
| 6 | 3.53% | EBook in CD-ROM format |

The answers of the question "Have you purchased any eBook or pay for browsing or assessing any eBook in the past?" were staggering: only 4.71% respondents have provided financial remuneration in order to receive or access electronic publication (Table 5). Most eBook readers are not eBook consumers, say have never spend any money on eBooks. Last question showed that 20.59% respondents got eBooks from websites that sale eBooks or charge in other ways. However most of them just accessed trials or free titles and few finally decided to purchase. The demand of eBook readers is far from the demand of eBook market.

Table 5: Question 3: Have you purchased any eBook or pay for browsing or assessing any eBook in the past?

| Frequency | Percentage | Option |
|---|---|---|
| 8 | 4.71% | Yes |
| 162 | 95.29% | No |

The data from question 2 and question 3 showed the long existing "free or not free" dilemma for digital content. This condition is prevalent for most digital content distributed in the internet like mp3, movie and so on. A similar survey about eBooks users in UK indicated that 38% eBook users had bought eBooks (Gunter, 2005). Although this proportion was still quite small, it's much bigger than the number in China.

## Reading Device for EBook

In Question 4, eBook users in the respondents were asked about the equipment they used to read eBooks. Respondents could indicate any equipment they used, which meant that some respondents could select more than one type of equipment. The great majority (90.59%) of the respondents replied reading eBooks on their desktop computer and 24.71 per cent readers used laptop computers. A little more than one in ten used personal digital device like Pocket PC and Palm (24.71%), and one in five used mobile phone (22.35%) (Table 6). There was no one used dedicated reading device to read eBook at all.

Table 6: Question 4: What type of device do you generally prefer to read eBook titles on? (Please check all that apply)

| Frequency | Percentage | Question |
|---|---|---|
| 154 | 90.59% | Desktop PC |
| 42 | 24.71% | Laptop |
| 21 | 12.35% | Personal Digital Device including Pocket PC, Palm |
| 38 | 22.35% | Mobile Phone |
| 0 | 0.00% | Dedicated Reading Device |
| 0 | 0.00% | Other (please specify)… |

This condition for reading devices was very prevalent in the eBook market of many countries in the world. A survey of electronic books users in the UK showed similar data: personal computer, especially desktop PC, was still the main equipment for eBook reading (91%) and very small proportions used dedicated e-book reader (7%) (Gunter, 2005). The EBook User Survey 2006 by the International Digital Publishing Forum showed that only 4% readers used dedicated reading device ("EBook User Survey 2006", 2006)

Because Chinese eBook market initiated later than developed countries, most Chinese eBook companies have learnt from foreign fellows and all focused on the *soft eBook* rather than the *hard eBook.* As I've mentioned above, there is only one eBook companies Founder APABI developed dedicated reading devices for their owned eBooks format among the all the eBook companies. Today it's extremely difficult to buy a specific hardware for eBook reading in Chinese epublishing market. Ray Kurzweil has said, "Inventing is about catching the wave. Most inventions fail not because the inventor can't get them to work but because the invention comes at the wrong time" (Riordan, 2003). Today the eBook companies try to avoid that epithet from being applied to itself, so most of them do not want to invest a lot to hardwires in the "wrong time".

Compared with dedicated reading device, mobile phone is a more popular device for eBook reading in China, the world's largest mobile phone market. According to the Ministry of Information Industry, the number of mobile phone subscribers has increased to 455 million in October 2006 ("October 2006 Monthly Statistic Report of Information Industry", 2006). Today more and more mobile phones have been equipped with the eBook reading function and become a better choice than dedicated reading device owing to their multi-functions.

## EBook Format

Today, in both global eBook market and Chinese eBook market there are various eBook formats in use: TXT, RTF, DOC, HTML, CHN, PDF, DJVU, etc. This condition can be referred to as the "Tower of e-Babel" behind which there are a number of issues and organizations involved in developing different eBook standards.

Question 5 provided a probe to the market share of eBook formats from end-users' perspective. Generally speaking, eBook formats can be classified into to two groups: open format and close format. The former is public to everyone and the latter is developed by eBook companies. PDF was ranked the first (34.71%) as the most popular eBook format by respondents, HTML, the basic format for World Wide Web, was ranked the second (31.76%), and the bundle of text-based formats, including TXT, RTF, and DOC, was ranked the third (27.65%) (Table 7). CHM, known as "Microsoft Compressed HTML Help, is a proprietary format based HTML". In CHM format, multiple pages and embedded graphics are distributed along with proprietary metadata are compressed a single file. If we consider both HTML and CHM as a bundle, the web-based formats have the largest share in the eBook market.

Table 7: Question 5: Which eBook format have you read? (Please check all that apply)

| Frequency | Percentage | Option |
|---|---|---|
| 47 | 27.65% | TXT, RTF, DOC |
| 54 | 31.76% | HTML |
| 35 | 20.59% | CHM |
| 59 | 34.71% | PDF |
| 27 | 15.88% | CEB, XEB (Founder APABI) |
| 43 | 25.29% | PDG (Superstar) |
| 26 | 15.29% | SEP (Sursen) |
| 39 | 22.94% | CAJ (Tsinghua Tongfang) |

By comparison, the eBook formats developed by eBook companies had smaller shares in respondents' choices. In this group, PDG format by Superstar was the most popular (25.29%), CAJ, the ejournal format by Tsinghua Tongfang, was ranked the second (22.94%), CEB and XEB by Founder APABI was ranked the third (15.88%), and SEP format by Sursen had the smallest share (15.29%).

The eBooks in open formats are generally created by individuals and free distributed in the internet. Among these eBooks, most of the modern titles are converted from paper books without the rights holders' authorizations, namely pirated. They are widely embraced by eBook readers because they are easy and free to obtain. In the contrast, the eBooks in close formats are generated by eBook companies who add DRM (Data Rights Management) to eBooks and sell them, and few people would like to pay for them. Because eBook companies also provide some titles for free trials, the shares of close formats eBooks are not so small in the end-users' reading experience though only 4.71 per cent readers have bought eBooks.

## Choices for eBook Genres

In Question 6, eBook users were asked, "What were the main genres for your eBooks?" The selections of respondents indicated that fiction publications (65.88%) and technical books (56.47%) were among those most popularly read (Table 8). Statistics indicated that information seeking and entertainment seeking were the top two aims for Chinese internet users ("A Statistical Report of the Development of the Internet in China (2006/1)", 2006).

It is not surprising that not matter readers from both east and west shared an universal taste. Fiction was the most selected theme. In the International Digital Publishing Forum's 2005 eBook Bestseller List, the most popular genres have been identified as science fiction and romance novels, non-reference textbooks, and 27 of all 30 bestsellers were fiction, most of which were science fictions. ("International Digital Publishing Forum's 2005 eBook Bestseller List", 2006). *The EBook User Survey 2006* by IDPF also showed the prevalent entertainment seeking trend in eBook readers.

Table 8: Question 6: What were the main genres for your eBooks? (Please check all that apply)

| Frequency | Percentage | Option |
|---|---|---|
| 96 | 56.47% | Technology about computer and programming |
| 112 | 65.88% | Fiction |
| 75 | 44.12% | Non-fiction Literature |
| 35 | 20.59% | Magazine |
| 46 | 27.06% | Academic Articles and Journals |
| 56 | 32.94% | Education and Textbooks |
| 34 | 20.00% | Ancient Literature |
| 69 | 40.59% | Other (please specify)… |

By comparison, the popularity of technology books about computer and programming are not only owing to the readers' taste but also the environment of internet. Most eBooks people accessed are free eBooks that created by individuals, and their genres are largely determined by the originators' choices. Most originators are some kind of computer "geeks" and their genres choices are related to hobbies and interests. In another hand, IT information is also one of the most wanted information for internet users. Considering the culture background of internet and the demand, it's no wonder to see the massive distribution of eBooks about IT technology.

# Electronic Books or Paper Books

To understand how users choose between electronic books and paper books, a couple of questions "why choose eBooks?" and "Why choose paper books?" were asked in this questionnaire. For the question 7, "EBooks are free or cost less" (65.88%) were ranked as the most frequent reason, and "EBooks are easy to access" was ranked the second (60.00%) (Table 9).

"EBooks save space" was ranked the third (45.88%), and "EBooks are searchable" was ranked the forth (44.71%) in question 7. The top four reasons to choose eBooks all indicated that readers selected electronic books mainly because of eBooks' technical advantages: costless to copy, easy to transfer, searchable and easy to store. Most respondents who selected "EBooks are searchable" in this question also selected "have got eBooks from websites of public libraries or libraries in universities or schools" in question 2. A respondent wrote "When I wrote my thesis, I found that searchable ejournals were more convenient for reference than paper documents." For the reference works, eBooks have the advantage of being easier to search and easier to annotate.

Table 9: Question 7: What were the main reasons for you to choose eBooks? (Please check all that apply)

| Frequency | Percentage | Option |
|---|---|---|
| 31 | 18.24% | No relevant paper book titles are available |
| 102 | 60.00% | EBooks are easy to access (more convenient than going to a bookstore) |
| 38 | 22.35% | EBooks are available to read anywhere |
| 35 | 20.59% | EBooks offer to access to new title |
| 78 | 45.88% | EBooks save space |
| 112 | 65.88% | EBooks are free or cost less |
| 32 | 18.82% | EBooks are good trials before buy paper books |
| 76 | 44.71% | EBooks are searchable |
| 9 | 5.29% | Foreign books can be easier to access in eBook format |
| 46 | 27.06% | Other reason (please specify)… |

Considering that few users really bought eBooks, "free" is obviously the most important reason to choose eBooks rather than paper books. Respondents' comments in this survey also showed that people who hold the "free" opinion to eBooks are quite prevalent:

"When a new book that I'm interested published, I would try to find a copy from internet just like downloading MP3."

"I don't think I will pay for eBooks. I just download free ones from internet."

"How can eBook sellers charge so much for eBooks! EBooks should be much cheaper than paper books."

From these comments, it is not difficult to find that electronic book is at a similar condition with most digital content in the internet like music, movie and so on. Because digital data can be easily copied and distributed, eBooks always face serious piracy problems. The data from Question 5 showed that a large proportion of eBooks was created by individuals and circulated free of charge over the Internet. From end-users perspective, they are easier to obtain compared with the eBook created and sold by eBook companies.

"Authorities may be well-defined or ill-defined ("A Framework for the Epublishing Ecology Public: Comment Draft Version 0.78", 2000)." Government has defined the law to protect the intellectual property, but the social norms may play more important role in the ecology of epublishing in which digital objects mainly flow through the internet. It's hardly possible to control the massive piracy of anonymous Individuals in the internet. According to the statistic by China Institute of Publishing Science, there are only 60.6 per cent people in China know the concept of copyright (Sampling Investigation Report on Reading and Buying Inclination of People Across China (2006), 2006). In such an environment, it is not surprising that end-users fully enjoy the advantages of eBooks and totally ignore the interest of the originators and rights holders. have more influence to the electronic market and it cannot be changed in short time.

For the question 8, "EBooks are hard to read and browse on computer screen" (71.76%) was intensively selected by respondents for choosing paper book rather than eBook. Many eBook users indicated that they did not like reading continuously from a computer screen. In respondents' comments, many of them indicated that cathode-ray tube (CRT) display makes their eyes tired after several pages reading and still feel more comfortable for reading the paper pages which cause less eye strain over extended reading time. The screen issue is regarded as a long existing problem that impedes users' acceptance to eBooks world-wide (Romano, 2001). Perhaps it will still be a universal problem for the eBooks before the huge amount of CRT monitors are replace by new display devices like liquid crystal display(LCD). In recent years, the LCD monitors are widely adapted by desktop PC and improve the experience for eBook reading, but and the bottleneck of computer screen still has not been solved yet.

Table 10: Question 8: What were the main reasons for you choose paper books rather eBooks? (Please check all that apply)

| Frequency | Percentage | Option |
| --- | --- | --- |
| 103 | 60.59% | No relevant eBook titles are available |
| 61 | 35.88% | EBook are difficult to find |
| 122 | 71.76% | EBooks are hard to read and browse on computer screen |
| 75 | 44.12% | EBooks have bad quality after digitized (mistakes in contents, graphic fonts…) |
| 68 | 40.00% | Other reason (please specify)… |

"No relevant eBook titles are available" (60.59%) was ranked second in question 8. Respondents also wrote comments about their disappointment for the content of eBooks:

"It's quite difficult to find latest titles in eBook version."

"Sometimes I can just get eBooks after the paper ones have been published for years."

"It's quit hard to find the latest releases by popular authors."

The success of eBooks will be still affected by the content like tradition books. After several years' growth, electronic publishing is still not able to be compared with traditional publishing in the aspect of content. Although there have been 148,000 titles of eBook published in China until May 2000, the amount of eBooks is still less than traditional books. Most of these eBooks were available only after its paper version had been published for years because of the difficulty for eBook Companies to get the authorization of new books from the authors who are less aware of electronic publishing or publishers who are afraid of eBooks and piracy to impact the sales of paper books. From the perspective of Intermediaries, the features of easy to transfer and copy has become both the advantage and disadvantage for eBooks because eBook can be much easier to pirated than paper book. The content issue is and will be another long existing problem for the development of eBook market.

From the answers of these two questions we can find that the technical pros and cons of eBook have become the most important factors that determine readers' choice between electronic books and paper books. In one hand, electronic books are more efficient than paper books from the various perspectives such as accessibility, transfer, delivery, search and storage. In another hand, electronic books are still not comparable to paper books in the display quality on CRT monitors which are the most popular display devices for desktop PC. Besides, the content is also an important factor that determined people's choice between electronic books and paper books.

# Conclusion

In China, the electronic publishing industry has gone through about ten years' development. This decade can be divided into three periods: the primitive stage, the trial and error phase and the developing period. The early stage was featured by the innovation of publishing technique, the boom of dotcoms based on the new conception of eBook and the high expectancy of vendor capital. The trial and error phase started with the winter of dotcom, continued with eBook companies' vicissitude for seeking the business model, and ended with the boom of B2B market of eBook. In the developing period, the whole eBook market was firmly shared by three companies, Beijing Founder APABI Technology Limited, Beijing Superstar Electric Company and Beijing Sursen Electronic Technology Company Limited, and they all mainly focused on B2B model owing to the Project 211 which greatly promoted the demand of eBooks in the education market.

Through several years' evolution, the electronic publishing market in China has grown, but the ecology of Chinese eBook market, the B2C market in particular, is still in infancy at present. The eBook industry faces many significant obstacles including copyrights issue, user acceptance and the reading device.

Copyright Issue is the most important problem for the electronic publishing industry in China. From the perspective from the competition, eBook Company should show the amounts of their eBooks as high as possible and so they did in front of their customers, but the public announced numbers were much smaller than the real

amount of eBooks from both Superstar and Sursen. However I still found the real numbers by searching the data from their universities customers who have bought digital libraries from these eBook companies and the result was staggering. Both Superstar and Sursen have announced that they have the authorization of the eBooks they hold from the authors, but the real number of the eBooks showed that it's impossible to pay for the authorization and that was also confirmed by my interviews. The prevalent copyright infringement has become a time bomb in the eBook industry, and these companies will have to pay for their ignoring to the right holders' interests in the epublication finally. Beijing Founder Electronics Co., Ltd. gets close relationship with most Chinese publishers due to its monopoly of the typesetting system in the traditional publishing market in China and has much stronger financial power to pay for the authorization, so Founder APABI is perhaps the only eBook company that has completely cleared copyrights issues for its eBook collections in China. But the amount of Founder APABI's eBooks is much less than either Superstar or Sursen.

The ill copyright awareness does not only exist in eBook companies but also in eBook end-users. There are a large proportion of eBooks are created by individual and distributed freely in the internet and they are more welcomed by readers than the eBooks sold by eBook companies. From the perspective of end-user, Question 2, question 3, question 5 and question 7 of this survey all indicated that eBook readers held negative opinion to *buy* eBooks. This condition is prevalent for most digital content in the internet like mp3, movie and so on. Due to the absence of law in the environment of the internet and the ignorance of copyright in the public, the eBook market is still in a rule-less condition.

Besides, the survey showed that readers have accepted electronic books but not the business model of eBooks companies today. The computer-based reading rate in the internet users is getting higher and higher, but the purchase rate of eBook stay almost at the bottom. More and more people read eBooks distributed in the internet because it is convenient to access and free to obtain. However eBook companies' B2C model which is copy with traditional publishing is somehow "against" readers' acceptance because people believe that anything in digital form should be abstained free of charge because it is "costless" for massive production. In a long time, it will be still difficult for internet users to pay for digital content like eBook. Easy to copy, one of the most important technical advantages of eBook, has become the biggest obstacle to its development in commerce. The success or failure of eBook industrial will depend not only on the change of users' attitude but also on the eBook companies' abilities to create new business models.

As a product of technology innovation in publishing, electronic book has showed its advantages over traditional media on modern reading. Electronic book is indeed more advanced than paper book but not in all the aspects. The questionnaire indicated that computer, mainly desktop PC, was the primary device of eBook readers, and this statue will not change in a long period. The CRT displays, which cause vision fatigue after extensively reading, perhaps bring the worst reading experience for eBooks, and that is definitely one of the most important obstacles for reader to widely accept eBooks. Today LCD display devices are becoming cheaper, and laptops, mobile phones, and other mobile devices that can be used for reading eBooks are getting popular, but there are still a large amount of CRT monitors in use. In another hand, the dedicated reading devices are extremely far from eBook readers in China today. The reading device issue will not be resolved in recent years until new display devices which can bring similar reading experience like paper and do not cost a lot.

# Bibliography

Encyclopedia of China Publishing House v. Beijing Superstar Electric Company, Civil Order No. 6243 (Beijing No.1 Intermediate People's Court 2002).

Encyclopedia of China Publishing House v. Beijing Sursen Electronic Technology Company Limited, Civil Order No. 11775 (Beijing Haidian District People's Court 2004).

Li Mingde v. Beijing Sursen Electronic Technology Company Limited, Civil Judgement No. 12507 (Beijing Haidian District People's Court 2004).

Li Shunde v. Beijing Sursen Electronic Technology Company Limited, Civil Judgement No. 12503 (Beijing Haidian District People's Court 2004).

Tang Guangliang v. Beijing Sursen Electronic Technology Company Limited, Civil Judgement No. 12508 (Beijing Haidian District People's Court 2004).

Xu Jiali v. Beijing Sursen Electronic Technology Company Limited, Civil Judgement No. 12506 (Beijing Haidian District People's Court 2004).

Zhang Yurui v. Beijing Sursen Electronic Technology Company Limited, Civil Judgement No. 12504 (Beijing Haidian District People's Court 2004).

Zheng Sicheng v. Beijing Sursen Electronic Technology Company Limited, Civil Judgement No. 12509 (Beijing Haidian District People's Court 2004).

Zhou Lin v. Beijing Sursen Electronic Technology Company Limited, Civil Judgement No. 12505 (Beijing Haidian District People's Court 2004).

Cultural Relics Publishing House v. Beijing Sursen Electronic Technology Company Limited, Civil Order No. 8717 (Beijing Haidian District People's Court 2005).

He Haiqun and etc. v. Beijing Sursen Electronic Technology Company Limited, Civil Order No. 7474 (Beijing Haidian District People's Court 2005).

Jin Chunming v. Beijing Sursen Electronic Technology Company Limited, Civil Order No. 24847 (Beijing Haidian District People's Court 2005).

Li Haiwen v. Beijing Superstar Electric Company, Civil Order No. 10909 (Beijing Haidian District People's Court 2005).

Tie Zhuwei v. Beijing Sursen Electronic Technology Company Limited, Civil Order No. 24845 (Beijing Haidian District People's Court 2005).

Beijing Normal University Press v. Beijing Superstar Electric Company, Civil Order No. 3741 (Beijing Haidian District People's Court 2006).

Li Zheying v. Beijing Superstar Electric Company, Civil Order No. 3613 (Beijing Haidian District People's Court 2006).

Liu Heping v. Beijing Superstar Electric Company, Civil Order No. 3614 (Beijing Haidian District People's Court 2006).

Shen Hongxin v. Beijing Superstar Electric Company, Civil Order No. 24971 (Beijing Haidian District People's Court 2006).

Social Sciences Academic Press v. Beijing Superstar Electric Company, Civil Order No. 28347 (Beijing Haidian District People's Court 2006).

Wang Xiaoming v. Beijing Superstar Electric Company, Civil Order No. 3610 (Beijing Haidian District People's Court 2006).

Zhang Yafei v. Beijing Superstar Electric Company, Civil Order No. 17712 (Beijing Haidian District People's Court 2006).

Zheng Hong and etc. v. Beijing Superstar Electric Company, Civil Order No. 3611 (Beijing Haidian District People's Court 2006).

Zhuang Yongping v. Beijing Superstar Electric Company, Civil Order No. 24972 (Beijing Haidian District People's Court 2006).

China Institute of Publishing Science. (2006). *Sampling Investigation Report on Reading and Buying Inclination of People Across China (2006).* China Institute of Publishing Science.

China Internet Network Information Center. (2006, January 17). *A Statistical Report of the Development of the Internet in China (2006/1).* Retrieved September 1, 2006 from China Internet Network Information Center: http://www.cnnic.net.cn/images/2006/download/2006011701.pdf.

Dipert, B. (2003). "Has Paper's Time Passed?". *EDN Europe, Vol. 48* (No. 9), p. 26.

Founder Holdings Limited. (2005, December). *2005 Annual Report.* Retrieved September 1, 2006 from Founder Holdings Limited: http://hk.founder.com.cn/upload/nianbao/car2005.pdf.

Gates, B. (2001). "E-books". *Executive Excellence, Vol. 18* (No. 4), p. 17.

Gunter, B. (2005). "Electronic Books: a Survey of Users in the UK". *Aslib Proceedings, Vol. 57* (No. 6), 513-522.

Hao, Z. (2005). *2004-2005 Annual Report of Publishing Industry in China.* China Book Press.

Hawkins, D. (2002). "Electronic Books: Report of Their Death Have Been Exaggerated". *Online, Vol. 26*, p. 42-48.

Herther, N. K. (2005). "The E-book Industry Today: a Bumpy Road Bbecomes an Evolutionary Path to Market Maturity. *The Electronic Library, Vol. 23* (No. 1), p. 45-53.

International Digital Publishing Forum. (2006, February). *EBook User Survey 2006.* Retrieved September 1, 2006 from International Digital Publishing Forum: http://www.idpf.org/doc_library/ surveys/IDPF_eBook_User_Survey_2006.pdf.

International Digital Publishing Forum. (2006). *International Digital Publishing Forum's 2005 eBook Bestseller List.* Retrieved September, 2006 from International Digital Publishing Forum: http://www.idpf.org/bestseller/bestsellers.htm.

Kang, K. (2001). "Bo Ku Wang: Feng Guang Bu Zai Dao Er Bu Bi". *China Reading Weekly, 20 June*.

Ministry of Education of the People's Republics of China. (n.d.). *a Brief Introduction of Project 211.* Retrieved September 1, 2006 from Ministry of Education of the

People's Republics of China:

http://www.moe.edu.cn/edoas/website18/info5607.htm.

Ministry of Information Industry of the People's Republic of China. (2006, December 22). *October 2006 Monthly Statistic Report of Information Industry.* Retrieved December 31, 2006 from Ministry of Information Industry of the People's Republic of China: http://www.mii.gov.cn/art/2006/12/22/art_27_27537.html.

Open eBook Forum. (2000, September 25). *A Framework for the Epublishing Ecology Public: Comment Draft Version 0.78.* Retrieved September 1, 2006 from International Digital Publishing Forum: http://www.idpf.org/doc_library/ecology.htm.

Ren, W. (2005). *2005 China EBook Annual Conference.* Retrieved September 1, 2006 from http://intro.apabi.com/20030404/about_2005_1.htm.

Riordan, T. (2003). "Patents: Investor Creates Software That Can Turn a Computer into a Cyber Poet". *New York Times, 24 November*, p. c7.

Romano, F. (2001). *E-books and the Challenge of Preservation.* Retrieved May 7, 2006 from www.clir.org: www.clir.org/pubs/reports/pub106/ebooks.html.

# Appendix:

## Questionnaire for Electronic Books Users

1. Have you read any eBook in the past?

- ❑ Yes

- ❑ No

2. Have you purchased any eBook or pay for browsing or reading any eBook in the past?

- ❑ Yes

- ❑ No

3. Where did you get eBooks?

- ❑ Websites that provide free eBooks to download or browse

- ❑ Websites that sale eBooks or charge in other ways

- ❑ Websites of public libraries or libraries in universities or schools

- ❑ EBook in CD-ROM format

4. What type of device do you generally prefer to read eBook titles on? (Please check all that apply)

- ❑ Desktop PC

- ❑ Laptop

- ❑ Personal Digital Device including Pocket PC, Palm

- ❑ Mobile Phone

- [ ] Dedicated Reading Device
- [ ] Other (please specify)…

5. In what format have you read e-Books? (Please check all that apply)

- [ ] TXT, RTF, DOC
- [ ] HTML
- [ ] CHM
- [ ] PDF
- [ ] CEB, XEB (Founder APABI)
- [ ] PDG (Superstar)
- [ ] SEP (Sursen)
- [ ] CAJ (Tsinghua Tongfang)

6. What were the main themes for your eBooks? (Please check all that apply)

- [ ] Technology about computer and programming
- [ ] Fiction
- [ ] Non-fiction
- [ ] Magazine
- [ ] Academic Articles and Journals
- [ ] Education and Textbooks
- [ ] Ancient Literature
- [ ] Other (please specify)…

7. What were the main reasons for you to choose eBooks rather than paper books? (Please check all that apply)

        No relevant paper book titles are available

- EBooks are easy to access (more convenient than going to a bookstore)
- EBooks are available to read anywhere
- EBooks offer t access to new title
- EBooks save space
- EBooks are free or cost less
- EBooks are good trials before buy paper books
- EBooks are searchable
- Foreign books can be easier to access in eBook format
- Other reason (please specify)…

8. What were the main reasons for you choose paper books rather than eBooks? (Please check all that apply)

- No relevant eBook titles are available
- EBook are difficult to find
- EBooks are hard to read and browse
- EBooks have bad quality after digitized (mistakes in contents, graphic fonts…)
- Other reason (please specify)…

9. Please write any comments you may have on eBook below:

Thank you very much!

# 电子书读者调查问卷

1. 您是否读过电子书 eBook（包括提供在线写作说的网站）？

   - ❏ 读过

   - ❏ 没读过

2. 您是否曾经购买过电子书？

   - ❏ 是

   - ❏ 否

3. 您从那里获得的电子书？

   - ❏ 提供电子书免费下载的网站

   - ❏ 提供电子书收费下载的网站

   - ❏ 公共图书馆的网站或大学图书馆

   - ❏ 电子书光盘

4. 通常您在下列哪种设备上读电子书？

   - ❏ 台式电脑

   - ❏ 笔记本电脑

   - ❏ PDA 等掌上电脑

   - ❏ 手机

   - ❏ 手持阅读器

   - ❏ 其他…

5. 您阅读过那种格式的电子书？（可多选）

   - ❏ TXT, RTF, DOC

- HTML
- CHM
- PDF
- CEB, XEB (方正 APABI 电子图书)
- PDG (超星电子图书)
- SEP (书生电子图书)
- CAJ (清华同方电子图书)
- 其他…

6. 您阅读的电子书的主题包括哪些？（可多选）

- 计算机、编程类图书
- 小说
- 非小说类畅销书
- 杂志
- 学术期刊著作
- 教材教辅类
- 古代典籍
- 其他…

7. 您会选择电子书的主要原因？

- 纸书没有相应的内容
- 电子书更方便获得
- 电子书可以随时阅读

- 当前最热门的图书出了电子书

- 电子书节省空间

- 电子书可免费下载或比纸书更便宜

- 可以先试读电子书再买纸书

- 电子书可检索

- 海外图书的电子版更容易获得

- 其他…

8. 您会选择纸质图书而不选择电子书的主要原因？

- 电子书没有相应的内容

- 电子书不好找

- 电子书不便于阅读

- 经电子化后书的质量下降（内容出错，图形字体等）

- 其他…

9. 如果您有任何其他观点，请写在这里…

谢谢！